**Late Ordovician geographic patterns of extinction compared with simulations of astrophysical ionizing radiation damage**

Adrian L. Melott, Department of Physics and Astronomy, University of Kansas, Lawrence, Kansas 66045. Email: melott@ku.edu

Brian C. Thomas, Department of Physics and Astronomy, Washburn University, Topeka, Kansas 66621. Email: brian.thomas@washburn.edu

*Abstrac.t*- Based on the intensity and rates of various kinds of intense ionizing radiation events such as supernovae and gamma-ray bursts, it is likely that the Earth has been subjected to one or more events of potential mass extinction level intensity during the Phanerozoic. These induce changes in atmospheric chemistry so that the level of Solar ultraviolet-B radiation reaching the surface and near-surface waters may be approximately doubled for up to one decade. This UVB level is known from experiment to be more than enough to kill off many kinds of organisms, particularly phytoplankton. It could easily induce a crash of the photosynthetic-based food chain in the oceans. Certain regularities in the latitudinal distribution of damage are apparent in computational simulations of the atmospheric changes. We previously proposed that the late Ordovician extinction is a plausible candidate for a contribution from an ionizing radiation event, based on environmental selectivity in trilobites. In order to test a null hypothesis based on this proposal, we confront latitudinal differential extinction rates predicted from the simulations with data from a published analysis of latitudinal gradients in the Ordovician extinction. The pattern of UVB damage always shows a characteristic strong maximum at some latitude, with substantially lower intensity to the north and south of this maximum. We find that the pattern of damage predicted from our simulations is consistent with the data assuming a burst approximately over the South Pole, and no further north than -75°. We predict that any land mass (such as parts of north China, Laurentia, and New Guinea) which then lay north of the equator should be a refugium from the UVB effects, and show a different pattern of extinction in the "first strike" of the end-Ordovician extinction, if it were induced by such a radiation event. More information on extinction strength versus latitude will help test this hypothesis.

## Introduction

Our goal is to describe and test a causal agent which has been proposed for one extinction event. There is a wide range of potential causes for Terrestrial mass extinctions.  Some of them are external to the Earth, and include bolide impacts (as widely discussed for the K/T boundary) and radiation events.  Among radiation events, there are possible large Solar flares, nearby supernovae, gamma-ray bursts (GRBs), and others.  These have variable intensity, duration, and probability of occurrence, although some generalizations are possible in understanding their effects (Ejzak et al. 2007).  Here we focus on gamma-ray bursts (Thorsett 1995; Scalo and Wheeler 2002). These are the most remote and infrequent of events, but by virtue of their power, a threat approximately competitive with, for example, that of nearby supernovae.  A GRB of the most powerful type (Woosley and Bloom 2004) is thought to result from a supernova at the end of stellar evolution for very massive stars with high rotational speed.  Much of their energy is channeled into beams, or jets, which include very high energy electromagnetic energy, i.e. gamma-rays and X-rays.  It is a testament to the power of these events that they were first detected when 1969-70 results from monitoring satellites designed to detect nuclear explosions on the surface of the Earth were triggered by events far across the observable Universe.  The distance to the events was not known until the 1990's, but when it was, their power became apparent. There are several per day in the observable Universe.  There are other kinds of events which are also potentially damaging, such as so-called short bursts and Solar flares, but rate information is only now beginning to clarify how much threat is likely from such sources.

Based on the rate of these events in the Universe as a whole, it is possible to estimate the rate and distribution of likely distances to events which irradiate the Earth (Scalo and Wheeler 2002; Melott et al 2004; Thomas et al. 2005; Dermer and Holmes 2005). These estimates were made as follows: the average rate in the Universe as a whole is scaled to the density of blue light.  Blue light is associated with large, hot stars, the kind which are precursors of GRBs.  GRBs in galaxies other than our own are too far away to cause damage to the Earth. From the density of blue light in our own Galaxy, we can estimate the likely rate in our Galaxy, which is in approximately one per 100,000 years. The radiation is known to be beamed, and only those pointed at the Earth will contribute to extinction, except for a possible enhancement to the general cosmic ray background.

We have done detailed computations of the atmospheric effect of what is the nearest likely such event in the Phanerozoic, based on the idea that only for this time period do we have any possibility of detecting the effects within the fossil record. Such an event would irradiate the Earth's upper atmosphere with approximately 10 times the intensity of sunlight (for typically about 10 seconds), but all in high-energy radiation such as X-rays and gamma-rays, even though approximately 6000 light years away.  There have been arguments that the GRB rate in galaxies like ours may be lower than originally thought (Stanek et al. 2006) as well as counter-arguments (Savaglio 2008; Savaglio et al. 2009; Ioka and Meszaros 2008).  However, due to nonlinearity in the nature of atmospheric solutions (Thomas et al. 2005; Thomas and Melott 2006; Ejzak et al. 2007) the possible rate difference would not greatly reduce the expected amount of damage to the biota. Much of our analysis is based on our "standard" extinction-level irradiation model, a fluence of 100 $kJ/m^2$, which would be the predicted most intense event in the Phanerozoic.

The amount of damage is dependent upon the intensity of the radiation which declines with distance from the burst.  When we combine the correction factor for beaming, the disclike geometry of the Galaxy and the trends we found in our simulations, we found that an event of the intensity described in the previous paragraph is likely every few hundred Myr.  Furthermore, one can approximate that events of a given damage level (defined as a % ozone depletion) happen at a rate proportional to the inverse square of the damage level.  Events with half a given damage level are likely to come four times as often.  There are other kinds of ionizing events possible, which can do similar damage but are probably a lower threat.  Supernovae may be a comparable threat (Gehrels et al. 2003; Fields et al. 2004), but our simulations would provide only partial understanding of their consequences.  Work is in progress to expand the understanding of supernovae (e.g. Thomas et al. 2008).

We include here only the effect of electromagnetic radiation, but it is also possible that the burst may be accompanied by a similar burst of ultra-high energy cosmic rays: high-energy atomic nuclei, mostly protons (Dermer and Holmes 2006; Dermer 2007). This would imply an additional class of damage effects of comparable magnitude and longer duration, which we do not include. We have made a start at understanding and including the atmospheric effects of cosmic rays (Melott et al. 2008).

The photons which make up the X- and gamma-radiation do not reach the ground (Smith et al. 2004). They travel a short distance in the atmosphere before interacting with it.  Nearly all of the energy goes into breaking up atoms and molecules by ionization and dissociation. Still, a burst of perhaps 10s length with a strong component of blue and ultraviolet light at the ground might be potential for some blinding or light burning of exposed tissue.  The changes in atmospheric chemistry are substantial, however.  For a nitrogen-oxygen atmosphere the biggest effects are consequences of breaking the strong triple bond of the $N_2$ molecule.  This enables the synthesis of many kinds of molecules which normally have very low abundances in the terrestrial atmosphere.  The most important for this discussion will be NO and $NO_2$.

## Secondary Mechanisms of Impact on Phanerozoic Biota

The primary effects are a consequence of the modified atmospheric chemistry.  The primary mechanism of damage is Solar UVB radiation which is able to get through the damaged stratospheric ozone layer.  We will first discuss the two smaller effects which have been analyzed elsewhere (Melott et al. 2005; Thomas and Honeyman 2008).

$NO_2$ is a brownish gas, which will absorb some of the sunlight which would normally reach the surface of the Earth.  This is not a large effect for our standard models, of order 1%, but it is much more intense near the poles, where the gases tend to accumulate.  It is possible that this perturbation could provide an impetus to flip the climate into glaciation at a time when it is unstable to this change, but lacking a "push." It has been suggested that the late Ordovician (Herrmann et al. 2004; Royer 2006) was in such a state. However, this possibility needs testing with a coupled ice-sheet and atmospheric general circulation model with a radiation perturbation.  Data exists which could be used to specify the perturbation (Melott et al. 2005), but to date this has not been studied further.

Secondly, the removal of odd nitrogen from the atmosphere by rainout as nitric acid is a source of nitrate beyond the usual sources (lightning and nitrogen-fixing by organisms). This is one of the principal mechanisms for recovery of the atmosphere over a period of order 5 years. Again, this is only of order 0.1 g per square meter for events such as those we consider here (Melott. et al. 2005), about equivalent to one or two years of production by lightning (Hewitt and Jackson 2003).  Thus there is an "acid rain" effect, but this is probably insufficient to provide a stress (Thomas and Honeyman 2008). Most biota (prior to artificial fertilizer) are nitrate-starved, and this would have been particularly true for early terrestrial environments. The additional nitrate deposition, while small, might provide a short-term positive effect, particularly for environments such as ponds where the effect would be concentrated and longer-lasting. Due to the extreme solubility of nitrates, it seems unlikely that any detectable fossil enhancement would remain from possible episodes, although some signals from Solar flares and modest cosmic ray enhancements are detectable in ice for recent events (e.g. McCracken et al. 2001; Thomas et al. 2007).

Both these mechanisms so far seem incapable of mounting extinction-level stress for reasonable estimates of astrophysical radiation, though more work needs to be done on possible climactic effects of opacity.

## The Primary Mechanism--Stress from Enhanced UVB

The oxides of nitrogen NO and $NO_2$ act as catalysts for the destruction of ozone, $O_3$. They continue to destroy it for years, without being consumed in the reaction, and are themselves only slowly destroyed by a variety of processes, including rainout and the action of sunlight (Thomas et al. 2005). The biota lose their UVB protection when the stratospheric ozone is depleted. UVB is known to be damaging based on current research on ecosystems.  We are currently experiencing substantial stress on (for example) amphibians (Blaustein et al. 1994; Blaustein and Kiesecker 2002) and phytoplankton (e.g. Boucher and Prezelin 1996)  with about 3% global average ozone depletion, and comparable UVB increase.  Effects on amphibians include reduced egg survival, lowered disease resistance, and deformities.  The primary measured effect on phytoplankton is reduced primary food production.

However, loss of primary production is not the only problem from enhanced UVB. As noted by Arrigo et al. (2003), these include (1) changes in phytoplankton species composition due to differential photoprotection and repair mechanisms, (2) changes in phytoplankton population structure that could result in modifications of the marine food web, as well as altering patterns of nutrient utilization, and (3) deleterious effects on any life stage of heterotrophic organisms found in surface waters (e.g., fish eggs) or in shallow benthic environments. Smaller organisms (being essentially transparent) are most readily damaged (e.g. Boucher and Prezelin 1996; Boelen et al. 2002; Kouwenberg et al. 1999). However, larger organisms can also be damaged directly (Blaustein and Kiesecker 2002). There is experimental evidence that organisms fed on UVB-damaged plankton themselves suffer reduced reproductive success including fewer eggs and more deformities (Kouwenberg and Lantoine 2006). A meta-analysis from the literature on UVB effects (Bancroft et al. 2007) found an overall negative effect of UVB on both survival and growth that crossed life histories, trophic groups, habitats and experimental venues.

While evidence of bolide impacts is found on the Earth's surface, the importance of likely damage from ionizing photons resulting in ozone depletion and major increase in damaging solar UVB reaching the surface has been recognized for decades. Recent studies have examined atmospheric effects in particular, due to supernovae (Gehrels et al. 2003) and GRBs (Thomas and Melott 2006 and references therein). Astrophysical events can reasonably double the UVB flux at the surface. Chemical mechanisms have been described in our past papers and elsewhere.

While all UV and even blue visible may be damaging, it is UVB that is highly variable depending upon the state of the Earth's stratospheric ozone ($O_3$) shield. Biological molecules, particularly DNA, strongly absorb energy in the UVB band and are damaged by it. It may be mentioned that although the benefit of the ozone shield varies with the level of atmospheric oxygen, it is much less that a linear effect, so a 10% or 35% oxygen atmosphere would not have extremely different shielding properties than our present atmosphere. The recovery of the ozone layer to near-normal levels from a few seconds burst of radiation would typically take a decade (Thomas et al. 2005).

Direct damage of marine organisms from UVB will be worse on the surface of the ocean or in shallow waters for larger organisms or eggs, as the attenuation length (reduction by 1/e) of UVB in water is of order a meter (Boelen 1999). Variation of extinction intensity with water depth is one of the patterns that exist in the late Ordovician extinction, as discussed by Melott et al. (2004), and was a plausibility argument for a role for UVB in this extinction event. Further plausibility arguments can be inferred from some other characteristics of this extinction. Glaciation and regression are usually blamed for this event, and would not be expected to cause extinction before they got underway. High-resolution stratigraphy showed that the primary extinction of graptolites occurred *before* the cooling of the climate and the fall in sea level (Brenchley et al. 2003). If an impulsive ionizing event such as a GRB were related to triggering the glaciation, damage from the UVB effects would take place before the glaciation and would be most immediately severe in planktonic organisms (such as graptolites), which are directly exposed to sunlight.

The most obvious effect is reduction of primary food production, since exposure to sunlight implies exposure to UVB. This damage is not confined to the surface layer.

Boucher and Prezelin (1996) estimated a 6-12% effect on reduction of primary food production over the entire water column at the present era in the region under the Antarctic ozone hole. Measurable depression of food production extended to depths of 30 m in that study. This does not give a direct estimate of the size of the effect we are likely to see: while the Antarctic is the most depleted region, the global average depletion from a GRB may reach 30% rather than the average of 3% or so we have experienced recently. But it does demonstrate that major effects are not confined to the surface layer. Arrigo et al. (2003) estimate approximately 60% loss of primary production from phytoplankton in the top meter and 20% integrated over the water column in the Antarctic. Since projected UVB increases from major astrophysical ionizing events are much larger than those in evidence today, a major food chain crash is possible. A period of a few years is short geologically, but very long compared to the lifespans of typical plankton.

We have described the background and general nature of the kind of damage likely to result from ionizing radiation events, which apply generally to the time in which an oxygen-rich atmosphere has existed. We have made previous (Melott et al. 2004) plausibility arguments that the Ordovician extinction is a likely candidate event. The goal of this work is to go beyond plausibility arguments by testing a falsifiable claim.

### Testing the Hypothesis

Such events are very "clean." There is little or no physical residue; there are no known radioisotopes generated with lifetimes longer than of order a Myr, except possibly if a strong burst of cosmic rays is included. Consistent prediction of damage based on cosmic rays is a matter for future work. In the meantime, however, it may be possible to test the ionization burst hypothesis against predictions that can be made for patterns of extinction. Inductive reasoning is most common in paleontology and astronomy, but there may be a role for falsification as well. Our simulations produce results as a function of latitude, and there exist extinction rates as a function of latitude for the Ordovician extinction from the paleontological record (Krug and Patzkowsky 2007). From the simulations we can construct a latitude-to-latitude ratio of UVB damage predicted. Our null hypothesis was "There will be no agreement between latitudinal UVB damage ratios from these simulations and data from the Ordovician extinction." This hypothesis is falsifiable. To be specific, we will define consistency as a particular scenario producing results lying completely within the error bars on the paleontological data. If we can reproduce the ratios, the null hypothesis should be rejected.

Such latitude dependence results primarily from the damaging "odd nitrogen" compounds escaping photolysis at the poles, and surviving longer. Also, due to less sunlight, less ozone is synthesized at the poles. This is related to the present high-latitude "ozone holes" that have been described. On the other hand, there is more intense sunlight near the equator. Our results combine the effects of these trends. The kind of latitude dependence seen in our models should be a general feature of atmospheric ozone depletion patterns, though it might be less intense at times of extreme global warmth, since stratospheric Brewer-Dobson circulation (Cordero et al.

2007) might be reduced. Our models are not strongly sensitive to differences in assumed $O_2$ or $CO_2$ concentrations such as appropriate for the Ordovician.

We have undertaken a new analysis of a number of "damage maps" from our atmospheric simulations, described elsewhere (e.g. Thomas et al. 2005), and added more simulations for better latitude coverage. Essentially, we combine the increased atmospheric transparency to UVB, with the length of the day and the angle of the Sun, to compute how much damaging UVB per square meter per day reaches the surface of the Earth at a given latitude and time of year. The increase in UVB is then convolved with a biological weighting function to estimate DNA damage (Setlow 1974). We find that it varies with latitude and time of year, as might be imagined. Direct damage to DNA is of primary importance for nearly transparent single-celled organisms, and its action spectrum is not greatly different in the UVB band from the action spectrum for other kinds of biological damage.

The pattern of ozone depletion depends upon the latitude and time of year of the burst. Detailed results have been presented elsewhere (Thomas et al. 2005). However, some generalizations can be described: (1) Effects tend to stay within the hemisphere of the burst. That is, a burst will tend to concentrate the ozone depletion toward the pole in the given hemisphere, because stratospheric transport is poleward. Therefore, in order to account for an end-Ordovician event, we will have to assume the burst was somewhere over the southern hemisphere, since the fossil record documents a southern hemisphere extinction. This is broadly consistent with astronomy, since there are more Galactic stars visible from the southern hemisphere than the northern. (2) Burst effects are more intense if the burst takes place in the autumn of the given hemisphere. This is because the odd nitrogen compounds can survive longer in the absence of sunlight and cause more depletion of stratospheric ozone. We cannot know the season of the burst for an event in the remote past, so we will have to average over events simulated at all seasons. Fortunately, this has no great effect on the latitudinal ratios, our primary concern here. (3) The results combine the effects of the typically more severe ozone depletion at the poles with the reduced level of sunlight at the poles, effects which work in opposite directions. These effects combine to give a peak damage level somewhere from the equator to the poles. At which latitude this peak comes depends upon the latitude of the burst, which can be systematically explored with the simulations. This we can test for consistency with paleontological data.

By covering a range of southern hemisphere burst latitudes, we can cover all possible cases which might be consistent with the Ordovician event. This is a simple one-parameter family based on the latitude of the burst. One of them must fit the data in order to falsify the null hypothesis and for the GRB ionization scenario to remain viable.

## Confrontation of Simulations with Data

Our simulation predictions will be confronted with extinction intensities from the Ashgillian in Krug and Patzkowsky (2007). They provide sampling standardized extinction intensities for the continents of Laurentia, Baltica, and Avalonia in their Figure 8. We have estimated these numbers for the Ashgillian from their Figure and will plot these data points against the results of our computations. Any extinction burst should have taken place at the Hirnantian, representing the last third of the Ashgillian, so some unavoidable time-averaging is implied. That is, their data will include extinction from all causes in the Ashgillian, but we will only be able to predict extinction from the possible UVB event component. This component would, however, have been a major extinction mode for the first strike of the Ordovician extinction, if an ionization burst were the basis of this event. This time averaging is an unavoidable limitation of our procedure and the time resolution of the data. We use the Paleomap project (Scotese and McKerrow 1990, 1991; C. Scotese personal communication 2008) latitudes of centroids of the continents from the Hirnantian.

Longer duration exposure will certainly lead to greater effects on most organisms. However, for simple organisms, especially phytoplankton, durations of hours or a few days are sufficient for serious effects to be observed (Boucher and Prezelin 1996; Neale et al. 1998; Llabres and Agusti 2006; Kouwenberg et al. 1999). Damage values in our modeling vary over timescales on order of weeks to a month. Therefore, here we consider maximum values in time, since exposure to a given value for even a few weeks will have a significant impact on nearly all organisms of interest.

While we are calculating DNA damage from a known action spectrum, we are making the simplest possible assumption: that DNA damage (our proxy for biological damage) is proportional to UVB dose, and that in turn is taken to be proportional to extinction intensity. Data do not exist to test the latter assumption. In regard to the former, some support exits for this linear relationship (Madronich et al. 1994; Neale et al. 2001). Furthermore, it holds for other kinds of damage as well: Smith et al (1992) state that "UVB inhibition of photosynthesis increases linearly with increasing UVB dose." Still, our approach is an assumption and data do not exist to test it in general for high UVB doses, nor are they likely to exist in the near future given the apparent preliminary success of the Montreal Protocol, which set limits on ozone-destroying compounds. Laboratory experiments on higher UVB levels are now unlikely to be done.

Although this assumption was made up front, we see that our results would follow with more general assumptions. Note that the two southerly data points are equal within their error bars, and would be equal even within $1\sigma$ errors (half those plotted), and that the more northerly point has lower intensity than the other two. For this reason our results would select the south pole and only it, even it we assumed merely that extinction intensity increased with UVB, not necessarily in linear proportion.

A typical "decay time" for damage to cells is of order several days (Das et al. 2001). Algae show overwhelming mortality after 5 days of 10% enhanced UVB (Kouwenberg and Lantoine 2006), and we are considering possible ~100% enhancements. Since our computed DNA damage values do not change greatly on the timescale of a week, we may take the worst daily value to approximate a dose. In fact, since there is little change in daily UVB dosage on the timescale of weeks to a month, we are effectively taking the most intense weeks over the few years of ozone depletion as our standard. One week is longer than the typical lifespan of phytoplankton. Most metazoan populations would have significantly declined as the base of the food chain crashed (Roopnarine 2006, and citations therein), possibly to extinction in this the 5-10 year period of ozone recovery. There will be additional direct effects on higher-level organisms and their eggs or larvae (e.g. Bancroft et al. 2007). We will compare the data with this set of assumptions for estimating the trend in the rate of damage to the biota as a function of latitude.

In order for the comparison to be meaningful, it is important to understand trends in the simulations. Figure 1, then, contains the summary of our results for the "worst week" assumption. As discussed earlier, such a burst would have to be over the southern hemisphere. We therefore did simulations with burst over latitudes spaced 15° apart from the equator to the south pole. This plot shows the computed intensity of the biological effects as a grayscale over latitude, when the burst latitude is varied between 0° and -90°. As can be seen, damage intensity peaks in the tropics for most of the phase space, but this runs counter to trends in the Krug and Patzkowsky (2007) data, for which the lower extinction intensity is closest to the equator.

We can refine the comparison by plotting the simulations as lines together with the data. In order to prevent visual confusion, we only plot four lines, corresponding to bursts at -90, -60, -45, and 0 degrees latitude. The four lines shown represent the relative level of intensity of damage of a burst which took place over the burst latitudes given. Since we are studying the latitude dependence of extinction rates, we show a number of lines exploring the latitude dependence of extinction for a variety of possible burst latitudes. The level of damage on each line is normalized to the maximum value at the equator (actually the +5 to -5 band in the simulations) averaged over all simulations at all latitudes and burst seasons. The three data points represent the sampling-standardized Ashgillian extinction intensity estimated by eye, along with 95% confidence interval error bars, from Krug and Patzkowsky (2007), Figure 8.

Note that the normalization of the data with respect to the simulations is free, such that the set of data or the lines can be moved up or down as a whole. The question becomes, "When the normalization is chosen so that a line passes inside one set of error bars, does it lie within the other two?"

We have found that a south polar burst fits quite well. The others in Figure 2 obviously do not, as the trends are completely wrong. We have found that a burst at -75° lies just outside the error bars, with no amount of vertical slide bringing it into agreement. Therefore, we conclude that the data are consistent with a burst between -75° and -90° south latitude.

## Conclusions and Discussion

Our results can be said to be consistent with the data, in that South Polar bursts reasonably reproduce the trend of the data. Thus we reject the null hypothesis stated above. Note that this does not prove the idea of a role for ionizing radiation in the extinction, but does show that it can be consistent with the latitudinal pattern.

Although the end-Ordovician extinction is entirely synchronous with expected downturn in the 62 Myr biodiversity cycle (Rohde and Muller 2005; Melott 2008 and references therein) there is no known reason to think that GRB bursts or similar phenomena would be periodic with this timescale.

There are inevitable limitations to our comparison. As mentioned before, we assume linear proportionality between the DNA-damage fluence of UVB and the extinction intensity, while things may well be more complicated than this. We note that while it is not impossible to have two GRB strikes on the Earth within less than a Myr, particularly from within one starforming region, it is not expected. Therefore the regularities we have described should only apply to the first pulse of the two within the end-Ordovician extinction. Having a burst nearly simultaneous with glaciation could be a coincidence, but if we seek a causal connection there should be climate simulation to see whether lost solar radiation from a reasonable burst can initiate glaciation under late Ordovician conditions (Herrmann et al. 2004), as suggested by Melott et al. (2005). Such a climate study has not been done, but data for initial conditions is available from the authors.

It is possible to use this result to make a new prediction. Certain paleogeographic reconstructions predict there may have been one or more northern hemisphere land masses, now part of northern China and New Guinea. We find that only a burst close to the South Pole works, so northern hemisphere atmosphere should have been largely free of any ozone depletion. Extinction rates, at least at the beginning of the extinction, should have been much lower there. The same should be true of the northern coast of Laurentia or any regions more than about 20°N latitude (see Figure 1). Of course, sea level effects associated with glaciation would certainly affect this area as well. Still, it may be possible to see some very different sequence of events, and this can possibly be tested against data. This region should definitely be a refugium from the UVB effects, especially since we are forced to assume a South Polar burst by the existing data. If these astrophysically-initiated UVB effects are an important fraction of the overall extinction effect, then northern hemisphere survivors might have served as a base for recolonization, beginning with Laurentia and parts of Australia. We note that Krug and Patzkowsky (2004; 2007) show that recovery was more rapid in Laurentia, as would be expected with invasion from its northern coast. (Most of the end-Ordovician fossil record comes from its southern coast). If there were latitudinal refugia in this extinction, recovery from this extinction should be unusually rapid, as found by Krug and Patzkowsky (2004) unlike what would be expected without such refugia. We note that the south pole also has low UVB effects due to the lessened sunlight, but the refugium effect there might be confused by the effects of glaciation.

Having initially made a plausibility argument for this scenario, we have now constructed and executed a test, which it has passed.  Knowing that only an extremely southern burst can work, we can formulate future tests based on this. These tests should include a look at the Ordovician extinction in the north coast of Laurentia, north China, and/or New Guinea and any other accessible Ordovician northern hemisphere regions, where the intial, UVB-mediated phase of extinction must be reduced if this scenario were correct.

There is always a tradeoff between quantity of data and its quality when controlled for a certain purpose. Partitioning existing data so that extinction from the first pulse of extinction is separate would help refine the test. Access to data more finely segregated by latitude, as suggested by Krug & Patzkowsky (2007), would also be useful.

### Acknowledgments

ALM acknowledges stimulating correspondence and conversations with M. Patzkowsky, paleolatitude information from C. Scotese, encouragement from J. Lipps and P. Ward to undertake this project, and helpful comments on the manuscript from Z. Krug, B. Lieberman, B. Anthony-Twarog, A. Miller, G. Hunt, M.J. Murray, and an anonymous referee.  The research was supported by supercomputing resources from the TeraGrid (NSF).

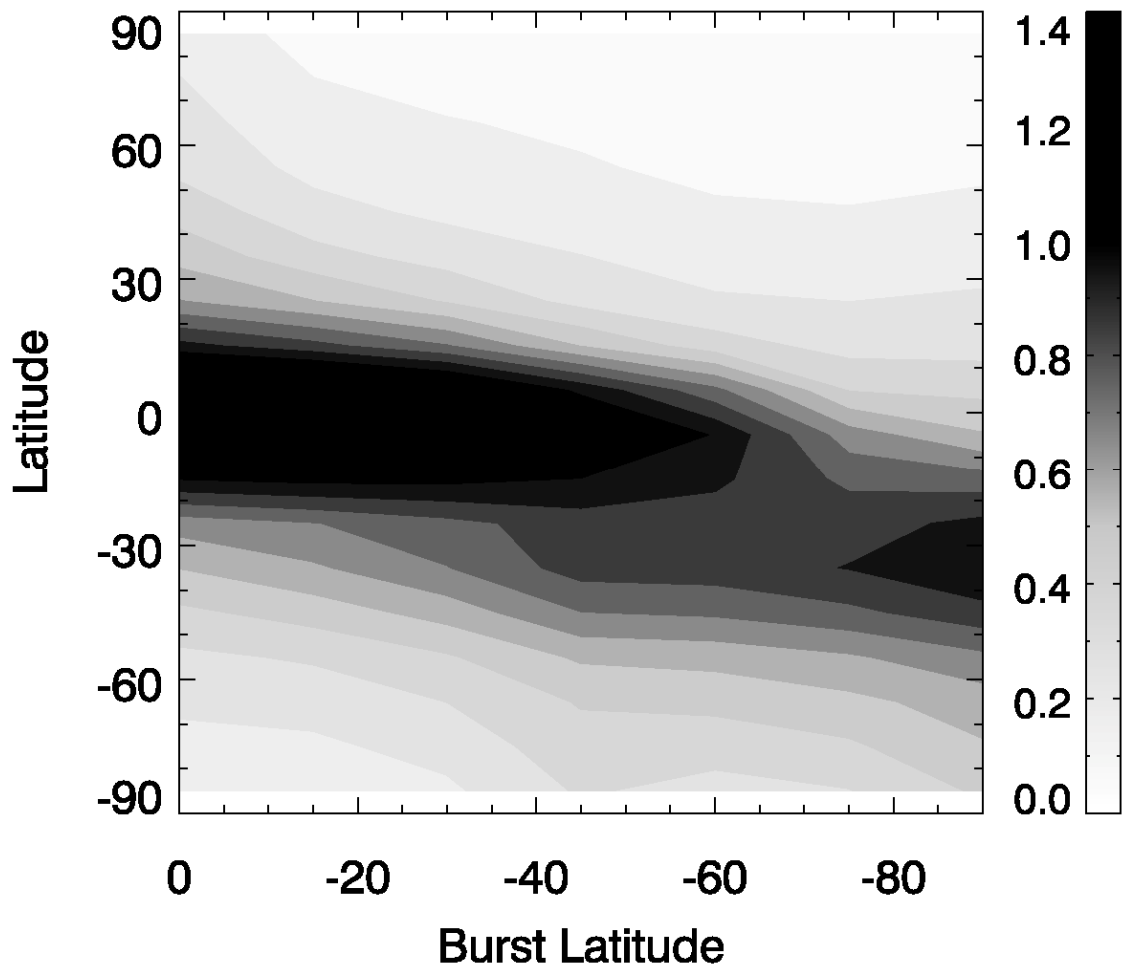

Figure 1: A summary of the maximum DNA damage intensity function from Solar UVB, based on computer simulations of ozone depletion from gamma-ray bursts over latitudes from the equator to the south pole, as found on the x-axis. The y-axis shows result latitudes. The greyscale represents maximum relative DNA damage intensities from UVB during the period of ozone depletion from gamma-ray bursts assumed to have taken place over four southern hemisphere latitudes, normalized to the average value of this maximum over all cases at the equator. As can be seen, for most burst latitudes, damage is greatest near the equator, where there is more direct sunlight. For circumpolar bursts, the ozone depletion moves so that mid-latitudes suffer the greatest damage. As we shall see, only near south polar bursts can fit the data, and consequently the northern hemisphere is largely safe from UVB effects.

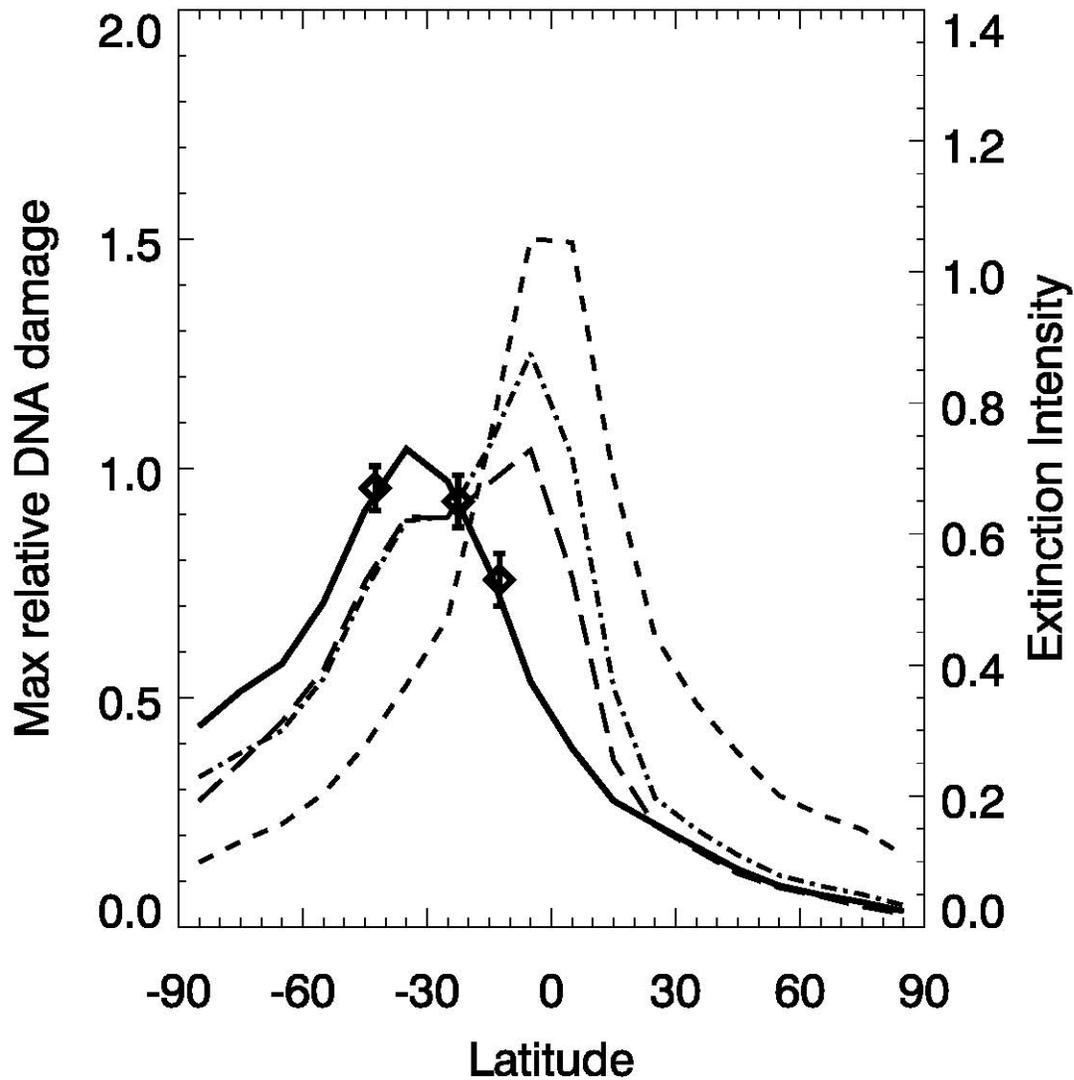

Figure 2: Computer simulations compared with data. The diamonds with (2σ) error bars represent sampling-standardized extinction intensities from Krug and Patzkowsky (2007), right hand axis. The lines represent maximum relative DNA damage intensities as in the grayscale of Figure 1. The lines correspond to bursts at latitude 0 degrees (short dash), -45 (dot-dash), -60 (long dash), and over the South Pole -90 (solid line). The south polar burst fits quite well, and a line corresponding to -75 degrees (not shown) cannot quite fit between the error bars of the data. Therefore, given our criteria, a burst south of -75 degrees latitude would fit the extinction data.